\documentclass[journal]{IEEEtran}
\usepackage{indentfirst}
\usepackage{caption}

\usepackage{cite}
\usepackage{url}

\usepackage{algorithm} 

\usepackage{algorithmic} 

\usepackage{multirow} 
\usepackage{enumerate}
\usepackage{amsmath} 

\usepackage{colortbl}
\usepackage[xcdraw]{xcolor}
\usepackage{array}
\usepackage{makecell}

\usepackage{booktabs}
\usepackage{hhline}

\ifCLASSINFOpdf
   \usepackage[pdftex]{graphicx}
   
   \usepackage{epstopdf}
\else
   \usepackage[dvips]{graphicx}
   
   \usepackage{epstopdf}
\fi

\usepackage{caption}
\begin{document}
\captionsetup[figure]{name={Fig.},labelsep=period}
\captionsetup[table]{name={TABLE.},labelsep=space}
%
\title{Intelligent Slicing of Radio Resource Control Layer for Cellular IoT:
Design and Implementation}
%
%
%

\author{Lian~Cao,
        Rongpeng~Li,
        Jon~Crowcroft,
		 Zhifeng~Zhao,
        and~Honggang~Zhang 
\thanks{Lian Cao and Honggang Zhang are with the College
of Information Science and Electronic Engineering, Zhejiang University,
Hangzhou 310027, China (e-mail: 21760203@zju.edu.cn, honggangzhang@zju.edu.cn).} 
\thanks{Rongpeng Li is with Zhejiang University, Hangzhou 310027, China as well as University of Cambridge, 15 JJ Thomson Avenue, Cambridge, UK CB3 0FD (lirongpeng@zju.edu.cn).}
\thanks{Jon Crowcroft is with University of Cambridge, 15 JJ Thomson Avenue, Cambridge, UK CB3 0FD (jon.crowcroft@cl.cam.ac.uk).}
\thanks{Zhifeng Zhao is with Zhejiang Lab, 
Hangzhou 310027, China (e-mail: zhaozf@zhejianglab.com).}}

\maketitle

\begin{abstract}
The cellular internet of things (CIoT) has become an important branch to cater various applications of IoT devices. Within CIoT, the radio resource control (RRC) layer is responsible for fundamental functionalities such as connection control and bearer establishment in radio access network (RAN). The emergence of various IoT scenarios and diversified service requirements have made both RAN slicing and intelligent control imperative requirement in RRC layer.
This paper focuses on enhancing standardized capabilities of CIoT RRC layer, by designing and implementing a new architecture which accommodate RRC slicing and intelligent controller. The architecture aims to realize functionalities of creating, modifying, and deleting slices in RRC layer, while the intelligent controller is added to satisfy various and dynamic service requirements of different IoT devices smartly. The proposed architecture is further implemented on an open-source software platform OpenAirInterface (OAI), on top of which the effectiveness of RRC slicing is validated and one proof-of-concept case to adopt reinforcement learning to dynamically tune discontinuous reception parameters therein is presented. Simulation results have demonstrated the effectiveness of the proposed intelligent RRC slicing architecture.

\end{abstract}


%
\IEEEpeerreviewmaketitle

\section{Introduction}
%
%
%
%
\IEEEPARstart{A}{s} the emerging of various IoT services, communication technology designed for  machines instead of humans has attracted a lot of attention. Since 2014, 3GPP began to study a new cellular system, namely cellular internet of things so as to better support ultra-low complexity and low throughput IoT devices\cite{tr45820}. So far, CIoT family have already contained three standards, that is, Extended Coverage Global System for Mobile Communications Internet of Things (EC-GSM-IoT), enhanced Machine Type Communications (eMTC) and Narrow Band Internet of Things (NB-IoT).

EC-GSM-IoT, which inherits from GSM and can be deployed by upgrading GSM network, satisfies the specific CIoT needs like extended coverage and reduced cost. Both NB-IoT and eMTC are modified and tailored from Long Term Evolution (LTE) technology, and can coexist with legacy LTE infrastructure, spectrum, and devices\cite{Soussi}. NB-IoT is specifically designed for low-power and low-throughput devices with 180kHz uplink and downlink bandwidth respectively, while eMTC has more bandwidth and mainly focuses on providing transmission for medium volume data services, such as voice service.

Unfortunately, the distinct requirements in terms of latency, bandwidth, power consumption implies that it is challenging to tackle them simultaneously using one single network. Thus, in addition to exploring the limits of network performance, CIoT network needs to be flexible and intelligent. In this regard, network slicing (NS) and artificial intelligent (AI) are regarded as promising enablers\cite{tr22864}.

\subsection{Related Works}
With the commercialization of NB-IoT and eMTC, there are still many shortcomings and challenges in current CIoT systems, especially in terms of power consumption and system flexibility. In this regard, it has been proven that energy consumption of IoT devices mainly comes from the transceiver\cite{power1, power2, power3}. Therefore, discontinuous reception (DRX) and its enchanced mechanism are adopted to reduce the power consumption by reducing the time spent on listening to the channel\cite{drx1, drx2}. Accordingly, there are also solutions trying to reduce power consumption by simplifying the signaling procedures\cite{simplesignaling} and designing the scheduling algorithm rationally\cite{schedule1, schedule2}.
Meanwhile, previous design of 3GPP system tends to provide a `one size fits all' system, which is only applicable to some specific services. However, future CIoT system is expected to be able to simultaneously provide optimized support for different configurations through various means. Flexibility and adaptability on network functionality and service are key features as the existence of various requirements. For example, vehicles and industrial devices may need low latency while others may not\cite{latency}, and some devices may require comparatively higher level security than others\cite{security}. For the moment, NS and AI solutions are incorporated into the cellular network to solve the aforementioned problems.

End-to-end network slicing, which aims to meet specific service requirements by carefully arranging essential functions and excluding redundant parts, implies to slice both the core network part and radio access network (RAN) part. 3GPP has already completed the specifications to support the service and operational requirements of network slicing\cite{ts22261}. Moreover, standardization of NS system architecture\cite{ts23501} and NS management and orchestration (MANO) capabilities\cite{ts28801} has also begun. Technically, a complete network slicing solution involves multiple aspects, including various virtualization technologies (such as abstraction and sharing of radio resources\cite{radiovirtualization,slicehua2019gan,sliceqi2019deep,sliceli2018deep}), lifecycle management for slices (e.g. 5G network slicing broker in \cite{nsbroker}). However, most of research on network slicing focuses on the core network, while sheds little attention on the RAN. The current discussion and research of core network slicing are based on the assumption that the RAN part has already motivated the needs of network functions and quality of service (QoS) required by different services simultaneously. In fact, existing radio access technologies (RAT) are independent of each other, providing different transmission support for different types of UEs, and they also have independent protocol stacks. For example, Cellular Vehicle To Everything (C-V2X) and NB-IoT are the symbolic technology of two main scenarios for 5G (i.e., URLLC and mMTC). They are deployed independent of each other, and operate in different frequency bands, with different RAN protocol designs. Therefore, it is necessary to coordinate heterogeneous resources and use network slicing to provision diversified user or service access, so that CIoT RATs such as NB-IoT can form different virtual networks in a logically single CIoT infrastructure and work as separate slices to provide services for different type of users. Correspondingly, a unified set of RAN customization and management standards need to be formulated in the access network as a whole. Besides, a flexible access network architecture should be further implemented to meet the functional requirements of RAN slices and provide adaptive radio resource management and control for diversified applications or services in 5G/6G.

In the field of RAN slicing, Ksentini \textit{et al}. referred to the design of software-defined networks (SDN) and added an eNodeB controller into the RAN which exchanges messages with the eNodeB through the southbound interface\cite{RANslice1}. From the perspective of wireless access protocol characteristics and configuration, Ferrus \textit{et al}. proposed to use slice configuration descriptors to manage the characteristics, policies, and radio resources\cite{RANslice2}. However, though the RAN slicing should solve the problems of radio resource abstraction and management and provide a flexible and configurable access network for various types of terminals in the coverage area, RAN slicing lags behind the research progress on implementing core network slicing and is still at an early stage of conceptual proposal and design. The technical trends of resource sharing and RAN as a service (RaaS) is unclear\cite{RANslice3}. Therefore, the implementation of RAN network slices for cellular IoT remains a challenge.

From the perspective of vendors and operators, the wide variety of network requirements, paired with a growing number of control parameters of modern RANs, has given rise to an overly complex system which is difficult to write maintenance, operation and fast-control software\cite{Ericsson}. There emerges an imperative need to both simplify the management and provisioning of different services and guarantee the quality of services.
In this regard, introducing AI to the RAN is promising to improve network performance and user experience and reduce complexity\cite{AIli2017intelligent}.
There are usually two types of architectures to implant AI for RAN, including AI on top of the RAN and AI embedded in the RAN\cite{Nokiareport}.
In addition to the aforementioned research, there have also been a large number of R$\&$D staff who invest in other key building blocks, including RAN interfaces, middleware, and machine learning platforms or toolkits.
Unfortunately, the RAN, which was almost void of any realistic AI simulation platforms, still has a long way to go towards intelligence.

\subsection{Contributions}
In this paper, we focus on the intelligent RRC slicing in CIoT, motivated by the fact that the RRC layer is responsible for implementing radio resource control protocol in the RAN, such as connection control between the UE and the network as well as management of UE states and contexts, and it impacts not only the implementation of the basic protocol functions defined in the standard, but also the network performance and user experience.

Briefly speaking, the contributions of this paper can be summarized as follows. 
\begin{enumerate}[1)]
\item We propose a flexible and agile RRC architecture for CIoT which incorporates the slicing inside an RRC layer and an intelligent controller on top.
\item The proposed architecture is successfully implemented and validated on the OpenAirInterface (OAI) platform, which is an open-source software platform for constructing communication systems and emulation of new architectures and technologies.
\item Inspired by studies which have demonstrated that RRC states and corresponding transmission mechanisms with different transceiver power have a very complex impact on network performance\cite{rrcstate1, rrcstate2}, we take the DRX mode as a proof-of-concept case to illustrate and verify the effectiveness of the proposed architecture.
\end{enumerate}

The remainder of this paper is organized as follows. In Section II, we present the purposed architecture for CIoT RRC slicing and describe the RRC slicing and intelligent controller in detail. In Section III, we provide an overview of the overall implementation. We take the DRX parameters optimization as a typical case and apply reinforcement learning algorithm for dynamic DRX parameter adaptation in Section IV, followed by a description of our experiments and results in Section V. Finally we conclude the paper and provide some insights for future work in Section VI.

\section{Intelligent RRC Architectural Framework for RAN Slicing}

In this section, we explain the design of RRC layer enforcing RAN slicing, and describe the details of the proposed intelligent RRC architecture for RAN slicing, as depicted in Fig. 1.

\begin{figure*}[!t]
\centering
\includegraphics[width=5.8in]{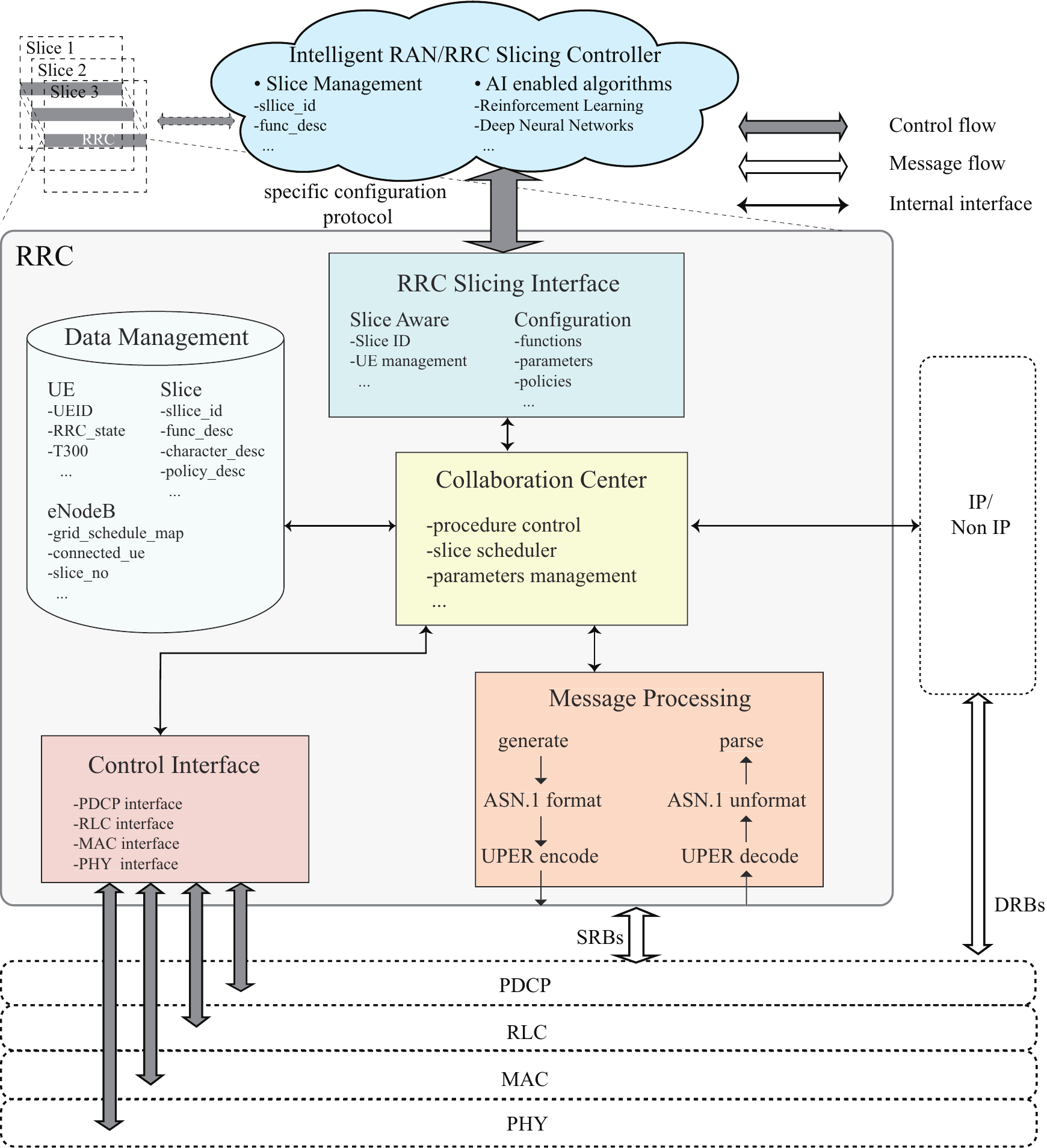}
\captionsetup[figure]{justification=centering}
\caption{Architectural framework for the realization of RRC slicing in CIoT network.}
\label{fig_arch}
\end{figure*}

It can be observed from Fig. 1, inside the RRC layer, slices are separately instantiated with each other and identified by unique slice identifiers which could be designed based on specific rules.
Furthermore, an intelligent RRC slicing controller is added to the framework to guarantee the intelligent arrangement of different slices. 
The intelligent controller can not only transmit slice configurations to RRC layer through the southbound interface but also control the RRC related parameters intelligently. 
In addition, a specific designed configuration protocol is defined in the southbound interface to transfer control messages between the slicing controller and RRC layer. The protocol should fulfill essential functions of adding, deleting, and modifying RRC slices, and be able to convey sufficient control information to ensure diversified functional control of different slices.
Notably, the slicing controller can be in charge of both RRC slices and slices of other  layers in the RAN. In other words, the intelligent RRC slicing controller can be extended to the much-broader RAN slicing easily by adding related interfaces to other layers.

\subsection{RRC Layer}
The RRC layer is located on the top of the access stratum of the control plane and is responsible for the control and management of various resources of the air interface, including transmission mode, logical channel, timers, constants, and etc., by means of different parameters\cite{ts36331}. The functions of the RRC layer consist of connection control, resource reservation, broadcast of system information, management of radio bearers, and security control of the access stratum. Most importantly, RRC layer realizes the above functions not only by itself but also through the configuration and management of the parameters to layers below it. In terms of the protocol, the control procedures of the RRC layer between UEs and base station (BS, e.g., eNodeB) is mainly achieved through a series of RRC messages, which consist of commands and configuration parameters for broadcast, multicast or unicast. The RRC layer entities of the BS and UE implement resources control to lower layers based on the information elements (IE) from the RRC messages. 

Based on the functional requirement of RRC protocol and links with slicing controller, the RRC layer contains the modules as shown in Fig. 1. 
\begin{itemize}
\item [-] The RRC mesage processing module is responsible for generating and parsing RRC messages, which are described according to the abstract syntax notation one (ASN.1) standard and encoded with unaligned packet encoding rule (UPER). 
\item [-] The collaboration management is a centralized management and scheduling module of the RRC layer. On one hand, it is in charge of all procedures in RRC layer, such as RRC connection setup. On the other hand, it interacts with other modules to realize slicing and parameters management. 
\item [-] The inter-layer control interface is designed for configuring the various protocol layers below the RRC layer, including packet data convergence protocol (PDCP) layer, radio link control (RLC) layer, medium access control (MAC) layer, and physical (PHY) layer. The configuration is implemented by transmitting control signaling inside the protocol stack like configuring the transmission mode of the RLC layer by informing acknowledged mode (AM), transparent mode (TM) or unacknowledged mode (UM).
\item [-] The data management module is specifically used to store the required contexts of all UEs in the network, and related network information of the BS such as configuration sets of network slicing. It can also record supplementary information such as resource usage status and traffic contexts in the BS.
\item [-] The RRC slicing interface in the layer is oriented to the RAN slicing controller and  responsible for performing management functions of the slices in RRC layer. As the research on network slicing in the field of RAN is still in its infancy, many contents have not yet been standardized, so the RRC slicing in this paper is based on a customized RAN slicing framework consisting of controller and actuator. The RRC slicing interface module plays the role of actuator in RRC layer, as shown in Fig. 1. It can meet the basic requirements of slice operations, including slice awareness, functions arrangement, QoS management, and isolation among slices at the RRC layer\cite{tr38801}.

\end{itemize}

\subsection{Intelligent RRC Slicing Controller}
The intelligent RRC slicing controller is located outside of RRC layer and may also be independent of the BS if necessary during the deployment. The controller is mainly responsible for the intelligent and flexible management of the slices, which could be RRC slices or even other layer's slices. It can also flexibly and dynamically configure different slices by accommodating interfaces for radio access network functions, management policies, and related parameters. The paper mainly focuses on the realization of creating, modifying and deleting slices with the aid of AI algorithms in the controller. Therefore, the northbound interface of the slicing controller and how the controller receive commands will not be discussed in this paper. The key purpose of the intelligent RRC slicing  controller can be summarized as:

\begin{itemize}
\item \textit{Flexibility}: Support flexible radio resource control among slices.
\item \textit{Scalability}: Easily add and delete the slices without influencing other slices.
\item \textit{Efficiency}: Appropriately manage the functions and parameters to make the slice efficient to both UE and the RAN.
\end{itemize}

In a word, the intelligent controller plays a key role in our proposed architecture by enabling flexible slices management and helping optimize the network and user performance.

\section{OAI based Implementation of RRC Slicing}

The implementation of the proposed architecture is further explained in this section. We establish the experimental prototype of our solution on the OAI platform, by modifying the current RRC design in OAI and enabling RRC slicing. We also reserve interfaces for implanting intelligent algorithms to help improve the effeciency of the architecture.
\subsection{OpenAirInterface}
OpenAirInterface is an open-source project for 4G/5G mobile communication systems that implements 3GPP technology and can run on general-purpose processors (i.e. servers and personal computers)\cite{OAI}. It provides a testbed for mobile communication systems when combined with off-the-shelf software defined radio devices like USRP series. Due to the implementation of comprehensive 3GPP protocols and multiplex simulation environments, it serves a solid basis on which we could build the prototype, validate our solutions and evaluate the performance.

\begin{table*}
	\caption{Implementation comparation between existing RRC architecture and our architecture}
    \centering
	\def\cg{\cellcolor{gray!20}}
\arrayrulewidth=1pt
\renewcommand{\arraystretch}{2}
\newcolumntype{C}{ >{\centering\arraybackslash} m{6cm} }
\newcolumntype{D}{ >{\centering\arraybackslash} m{3cm} }
\definecolor{Gray}{gray}{0.9}
\begin{tabular}{|>{\raggedright\arraybackslash}m{3cm}|m{6.8cm}|m{6.9cm}|}
\hline
   \multicolumn{1}{|>{\centering\arraybackslash}m{3cm}|}{\cg\textbf{Aspects}}
    & \multicolumn{1}{>{\centering\arraybackslash}m{6.8cm}|}{\cg\textbf{Existing RRC layer in OAI}} 
    & \multicolumn{1}{>{\centering\arraybackslash}m{6.9cm}|}{\cg\textbf{Our Prototype}}\\
\hline
       \multicolumn{1}{|>{\centering\arraybackslash}m{3cm}|}{{\cg}{RRC slicing}} & {Not support} & {Supported and slices running separately} \\
\hline
        \multicolumn{1}{|>{\centering\arraybackslash}m{3cm}|}{\cg{Modules}} & { Simple RRC layer integrated with other layers in the BS, which incorporates basic functions for one single RAT} & Including intelligent RAN/RRC slicing controller and module partition in RRC layer to enable flexible management for slices \\
\hline
		\multicolumn{1}{|>{\centering\arraybackslash}m{3cm}|}{{\cg} {Data storage}}& {Network contexts for UEs and eNodeB (Single RAT)} & {Network contexts and traffic contexts for UEs and slices, managed according to \textit{slice\underline{\space}id}} \\
\hline
        \multicolumn{1}{|>{\centering\arraybackslash}m{3cm}|}{{\cg} {Number of threads}} & { Single thread for one layer} &Multiple threads for different slices and flexible  management based on slicing configuration commands\\
\hline

        \multicolumn{1}{|>{\centering\arraybackslash}m{3cm}|}{{\cg} {RRC interface for management}}& {Fixed parameters, fixed policies and unable to modify when running} & {Capable to modify RRC parameters and policies by an intelligent controller}\\
\hline
        \multicolumn{1}{|>{\centering\arraybackslash}m{3cm}|}{{\cg} {Lower layer APIs}} & {Transfer parameters and messages, unable to distinguish between slices} & {Including \textit{slice\underline{\space}id} into all APIs to construct logical interfaces for different slices, enabling the interface to be customized based on slice}\\
\hline
		\multicolumn{1}{|>{\centering\arraybackslash}m{3cm}|}{{\cg} {Intelligent algorithms}} & {None } & {Reserving interface for implanting algorithms, easy for intelligent controller to configure RRC parameters dynamically}\\
\hline
    \end{tabular}
\end{table*}
The OAI platform is mainly programmed with C language and developed under the Linux system. It has implemented the core functions and protocols of eNodeB, UE, and EPC of LTE in accordance with 3GPP standards, and can construct mature and stable LTE system. In addition, the OAI platform also contains a large number of simulation environments and analysis tools for verification of various communication algorithms\cite{OAIdemo}. Currently, multiple organizations, institutions, and individuals are working together to implement 5G communication system with OAI.

The RAN slicing-oriented CIoT RRC layer implemented in this paper is developed and implemented based on the OAI platform. The operating system is the Ubuntu 14.04 release of the Linux which runs on an Intel Core i7-7700 processor (4 cores). Both the BS side and the UE side of the RRC layer are implemented on the platform. As other layers are not included in the architecture and slices of other layers or entities have not been completely realized in OAI, so our prototype only runs under the simulation environment \textit{oaisim}, but it is very close to the runtime situation with intact protocols and control functions.

\subsection{Prototype Implementation}
As the prototype of our architecture is built up on an OAI platform, NB-IoT is selected as one typical RAT of CIoT technologies. Considering the existing cellular IoT standards, the control plane (CP) and the user plane (UP) optimization scheme of NB-IoT EPS is different and cannot run at the same time in a device. Thus, they can barely be regarded as two different slices that can be implemented in software. In our solution, LTE technology, which is quite similar to LTE-M standard, can be regarded as another type of slice available in the CIoT slicing pool. The RRC functions of all three type of slices mentioned above are implemented in our prototype. 

As highlighted in Table \uppercase\expandafter{\romannumeral1}, our prototype is quite different from existing RRC architecture on the OAI platform. First, in our solution, we not only add a new entity named intelligent RRC slicing controller outside RRC layer, but also reorganized the RRC layer through module partition to enable the flexibility management of different slices. Therefore, it becomes accessible to control the parameters for different slices intelligently and specifically. Second, data stored in our prototype also include traffic contexts, and are managed according to the belonging slice, thus requiring slice-aware inter-layer interfaces. Third, to realize the RRC slicing in software, we use multiple threads in RRC layer and each represents a single task for one slice, so the slices run separately from each other and will have a minimum effects to others. Furthermore, through the application interface from intelligent RRC/RAN slicing controller, the RRC layer can receive configuration commands and manage RRC slices. Considering the importance of the whole life cycle, the identifier \textit{slice\underline{\space}id} is added into the intertask interface (ITTI) as a necessary parameter and make the original ITTI tool slice-aware. For each RRC slice in the BS, it has own task according to the RAT and configurations from intelligent RRC slicing controller. As a result, RRC slices operate separately and process their own messages. 
Fig. 2 provides an illustrative thread on how the BS could initialize the RRC slicing and create different slices. 

From the architecture design, it can been seen that all the modules need to be realized in OAI. And different from the traditional protocol realization on the platform, RAN slicing implies that all functions, parameters and procedures are slice-specific. In other words, each slice can have its own characters and orchestrate customized functions according to the RAN slicing controller.
\begin{figure*}
\centering
\includegraphics[width=7.13in]{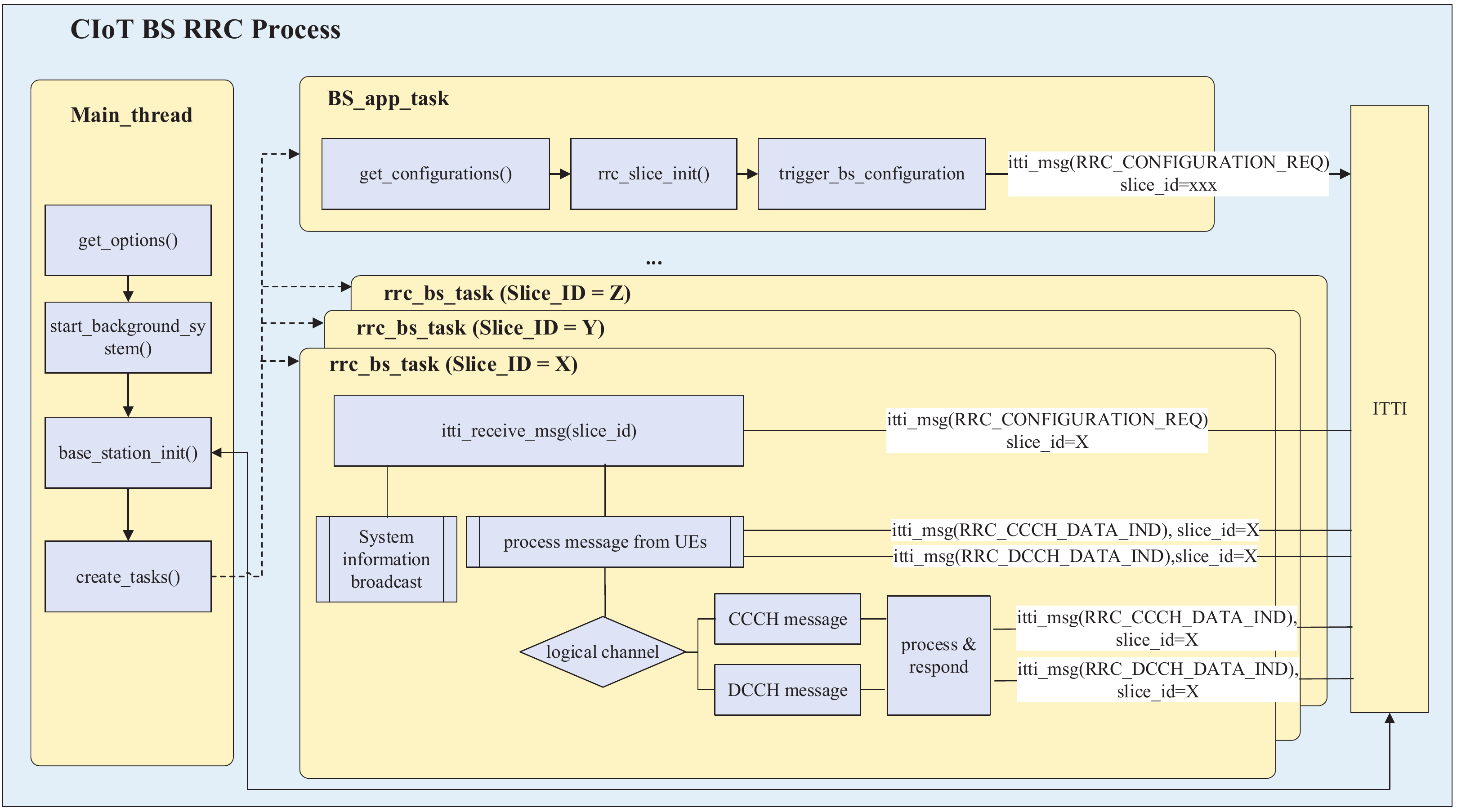}
\caption{RRC layer process implementing RRC slicing}
\label{2}
\end{figure*}

\subsection{Module Implementation Details}
In this part, the details of prototyping and realizing the modules in Fig. 1 is carefully provided.
\subsubsection{Message Processing}
\begin{figure*}
\centering
\includegraphics[width=5.8in]{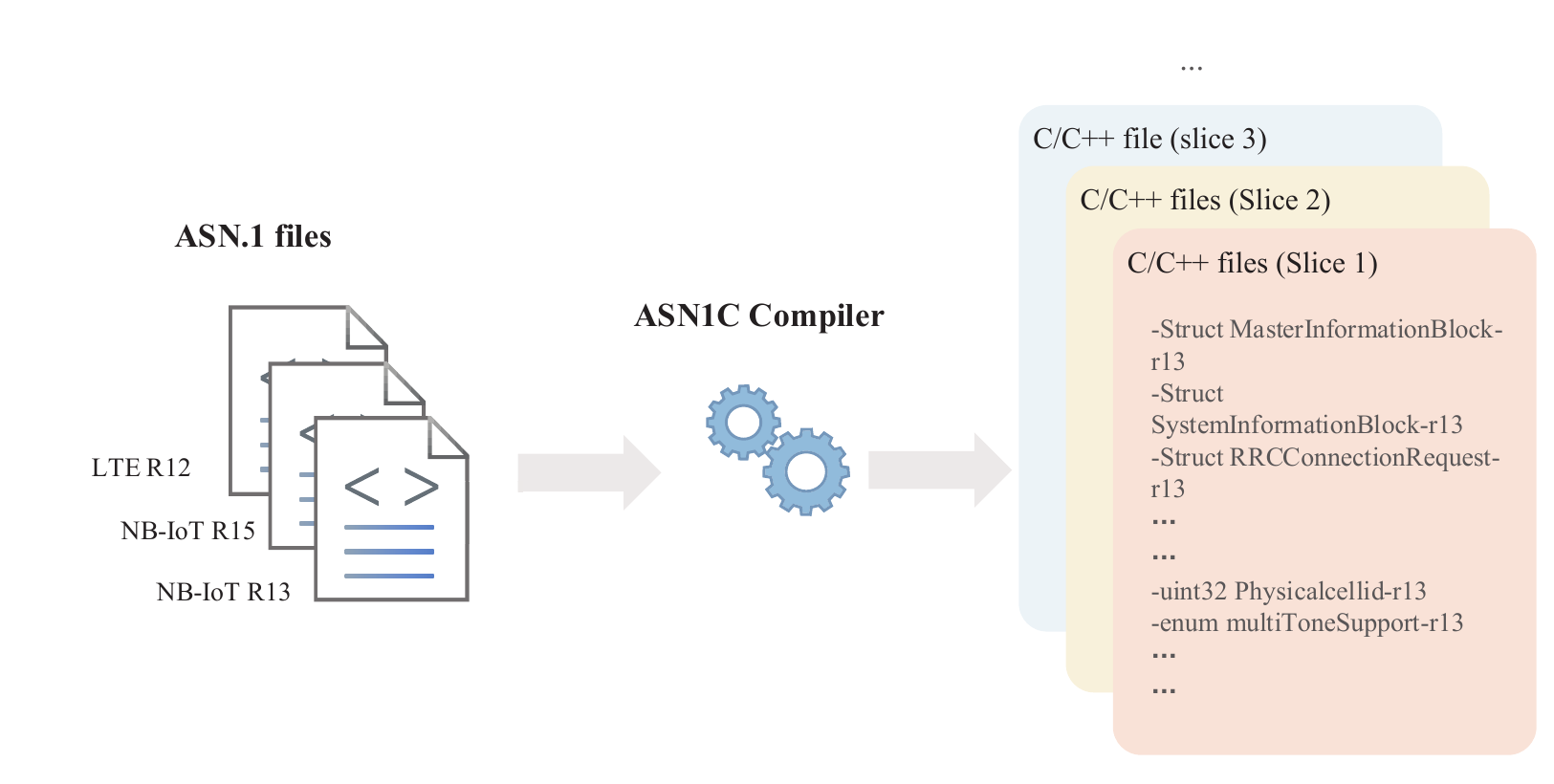}
\caption{Generation of slice-oriented RRC message data structure.}
\label{asn1c}
\end{figure*}
The RRC message processing is the core element of the RRC layer in CIoT system. The UEs and the BS exchange control information through RRC messages, which can only be parsed and understood at the RRC layer. No matter in CIoT networks or other cellular networks, RRC messages are represented by ASN.1 and encoded with UPER. 
Therefore, the message processing module mainly contains two functions, that is, using the ASN.1 programming language to represent RRC messages and UPER codec. Specifically, this module implements conversion between control plane data (such as context and other information) and UPER binary encoding from or to layers below. This module first generates the required RRC message in accordance with the RAT standard. Then the module sends the message to the corresponding logical channels and signalling radio bearer (SRB) for further transmission. At the same time, the module also parses and converts the received UPER-encoded message into a program-usable data structure. The PDU data that the message processing module outputs to or receives from lower layers is defined as \textit{logicalchannel\underline{\space}msg\underline{\space}RRC\underline{\space}PDU}, where \textit{logicalchannel} represents the name of logical channel through which the RRC message is transmitted. Grouping messages based on logical channels can facilitate lower-layer modules to handle different link layer operations, and also simplify the message parsing process of the RRC layer.
In this module, ASN.1 codec is implemented with the help of the tool ASN1C, an open source ASN.1 compiler\cite{ASN1C}. It can convert ASN.1 specification text into C language source code. The code generated by ASN1C consists of both type definition and codec function. The type definition is generated after the ASN.1 data structure initialization, and contains all the data types described in ASN.1. 
The ASN.1 data structure initialization scheme for RRC slices involved in this paper is shown in Fig. 3. Based on different RRC slices, the message processing module generates data structure files corresponding to the RRC messages of the current slice RAT during the initialization phase, right after that the command of adding a new slice is received. After the slice initialization is completed, the data structure prototype in the relevant source files can be accessed through the identifier \textit{slice\underline{\space}id} to implement the function of generating and processing the RRC message content.
The codec function can realize the conversion between this data type and the encoded data stream, i.e., encoding and decoding process. In addition to the ASN.1 compiler mentioned above, the codec function also involves a general runtime library that contains the basic rules for encoding and decoding ASN.1 primitives.
\subsubsection{Collaboration Management}
\begin{figure*}
\centering
\includegraphics[width=6in]{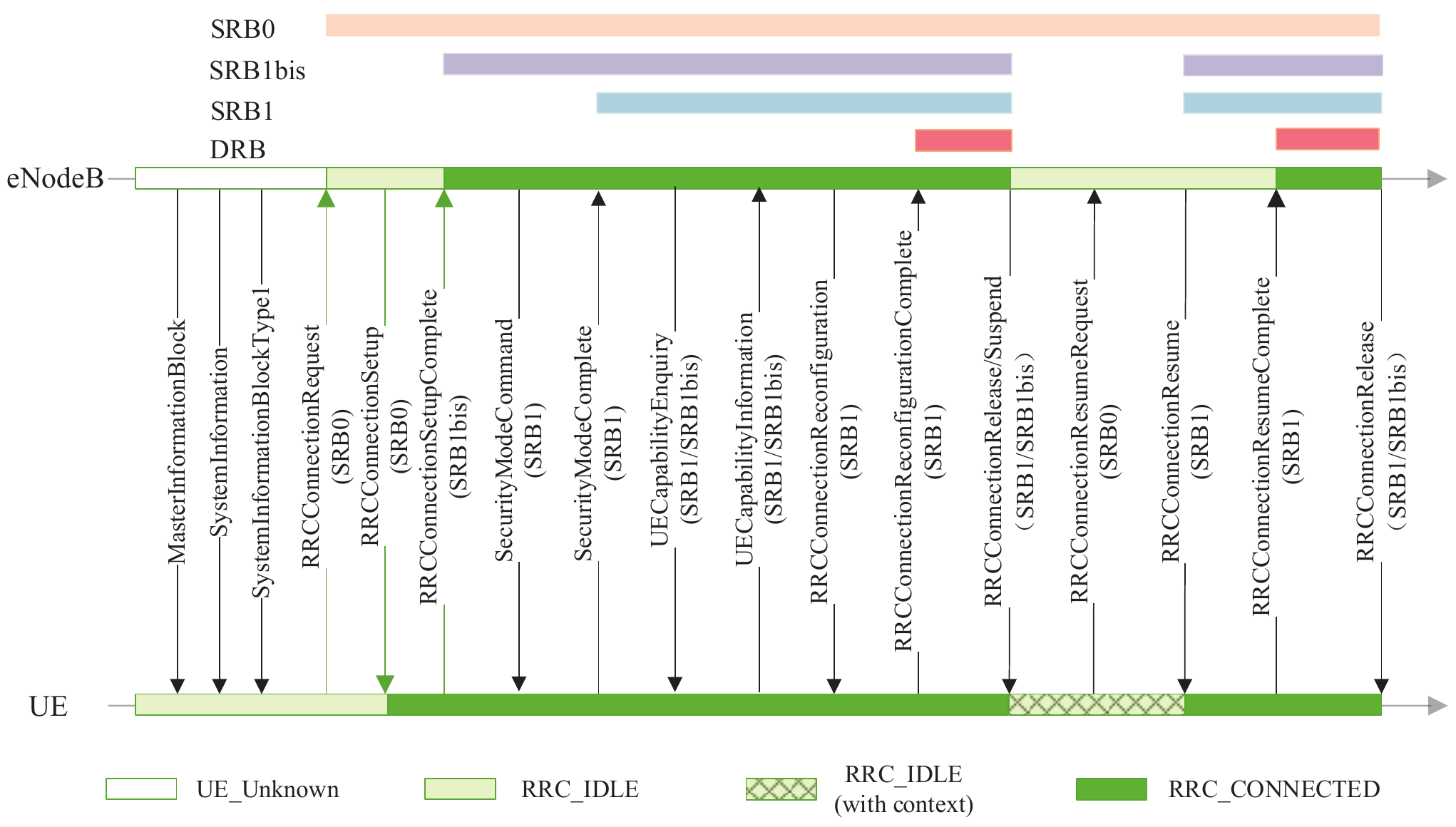}
\caption{RRC state and radio bearer flow with normal procedures in NB-IoT system.}
\label{RB and state}
\end{figure*}
The collaboration management module is very important in managing different tasks. 
On one hand, this module includes the thread pool management of all the main functions in the RRC layer, manages the communication between threads, and controls different slices according to the command received from RRC slicing interface. 
On the other hand, it is also responsible for the scheduling, execution and management of various RRC functions and other modules, and cooperatively manages the implementation of various contexts, procedures and state machine. As the execution of RRC procedures, the connection state machine and related bearers changes accordingly with the management of the module. An example of NB-IoT state and radio bearer flow under normal RRC procedures is shown in Fig. 4.
Procedure management is the main function of the RRC layer in CIoT and is implemented by the collaborative management module. Besides managing the RRC message transmission between the BS and the UEs, it is also responsible for the management of RRC status and contexts. RRC messages scheduled by the module are roughly divided into three categories: system information, paging and connection control messages. The system information broadcasting and paging procedures are initiated by the BS while UE is responsible for receiving and processing the indication information contained in the message. The connection control includes both uplink and downlink signalings. 
\subsubsection{Control Interface}
Through the definition of the inter-layer interface and the transfer of configuration parameters, communication between layers can be realized and RRC layer can achieve the purpose of controlling the functions of lower layers. From 3GPP TS 36.331, we can know that most of these inter-layer parameters involved the management and configuration of each layer are defined in \textit{RadioResourceConfigDedicated} IE which belongs to several specific RRC signallings. The particular content of this IE can be seen in Fig. 5. It contains many parameters related to power control, transmission reliability, and connection resource allocation, such as \textit{DRXCycle} in MAC layer and \textit{DiscardTimer} in PDCP layer, which represents the discontinous reception cycle and the maximum time to wait before discarding a PDCP packet respectively. As slice division is independent of each other in different protocol layers under the proposed framework, the interface of this module should transmit the signals with the \textit{slice\underline{\space}id} to make sure that lower-layer functions corresponding to the slice are configured consistently. In addition, because functions of different slices are isolated, the corresponding management and configuration parameter set of this module is also isolated based on the slice identification.
\begin{figure}
\centering
\includegraphics[scale=0.426]{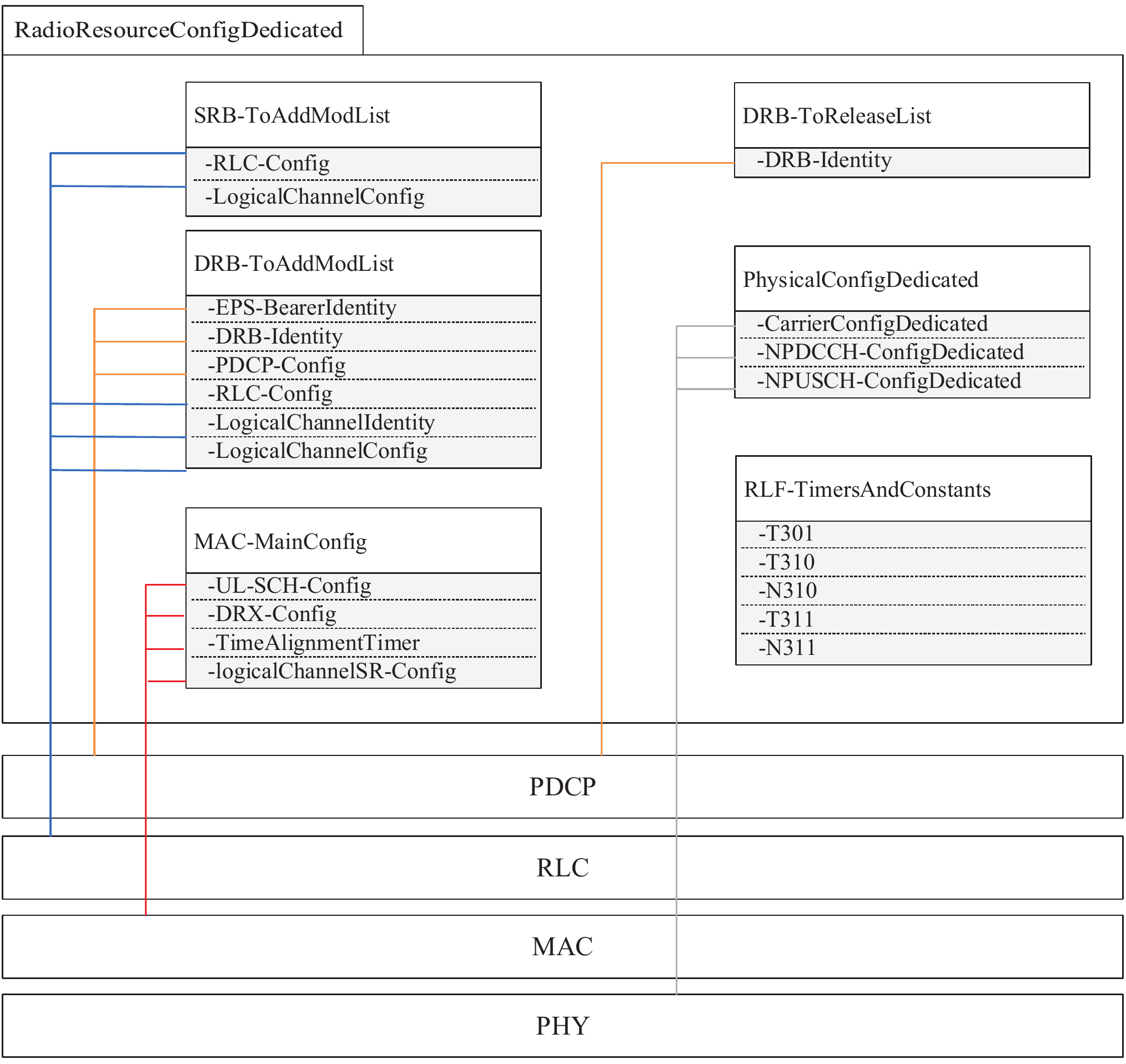}
\caption{Parameters in \textit{RadioResourceConfigDedicated} IE related with lower-layer configurations}
\label{RadioResourceConfigDedicated IE}
\end{figure}
\subsubsection{Data Management}
Based on the requirements of network slicing in the RAN, this prototype uses dynamic memory allocation to request memory space from the operating system in accordance with the slice. The stored contents include information about the UE's state machine, contexts, service requirements and so on. It's worth mentioning that data related to user traffic, such as UE's packet size, transmission time and historical configuration information, are useful for the optimization algorithms to improve the network performance but not required by functional protocol procedures. Therefore, they could be stored in a database\footnote{As 3GPP introduces a novel network entity named NWDAF (network data analysis functions), such a database can interact with NWDAF as well.} to record the traffic characteristics of the UE for further optimization.
Other modules can access the data management module to obtain the essential information, and other modules can also add, modify and delete the information in the data management module according to their authority. It is important to note that all data is managed according to the slice identifier \textit{slice\underline{\space}id}.
\subsubsection{RRC Slicing Interface}
When the intelligent RRC/RAN slicing controller sends a command to the RRC layer through this interface, the RRC slicing interface parses the command and completes corresponding functions such as slice establishment, configuration, maintenance, and modification based on the command. The format of the control command may be defined as a dedicated protocol with specially-designed descriptors, which are used to define the parameters, policies and functions of the slice. Once the slice defined by the intelligent RRC/RAN slicing controller is established, other modules in this layer also execute the corresponding processes and functions corresponding to the RRC slice based on \textit{slice\underline{\space}id}, so as to achieve the mutual isolation.
From the perspective of RRC slicing, the southbound interface of the slicing controller is simplified in the implementation, rather than based on a series of rigorous and unified control signals. Therefore, in our prototype, simply-defined control commands will be transmitted between the RRC slicing controller and RRC layer with customized functions and attribute descriptors to meet the requirements of managing multiple RRC slices logically. In our solution, each specific RRC slice manifests itself as a single thread running independently with each other.
\subsubsection{Intelligent RRC Slicing Controller}
The module is used as a smart manager to slices and can receive policies from the northbound interface. However, in our prototype, we simplify the implementation of the northbound interface and only realize the controller as an entity which can send commands to the RRC layer. We also add primitives to control RRC parameters and functions in this module to enable the flexible transplant of intelligent algorithms. Due to the separation of RRC layer, the algorithms will not impact the normal protocol flows except functions and parameters configuration. It is also slice-specific like other RRC modules. Our prototype only incorporates the controller for RRC slicing and intelligent control, but it can be extended to other layers easily.

\section{Intelligent Algorithm for DRX in CIoT system}
As depicted in Fig. 1, the RAN slicing controller is designed with APIs reserved for implanting algorithms which could intelligently control slices and accordingly influence the performance of network and UEs. As mentioned before, RRC layer can affect many aspects of connections between the UE and the BS by adjusting functions and timers. Thus, it is essential for intelligent RAN slicing controller to control the slices with related algorithms. In this section, we take account of the problem of energy efficiency and downlink transmission latency optimization under DRX mode as an example, so as to illustrate the effects of implementing reinforcement learning in RRC slicing controller and demonstrate how the reinforcement learning algorithm can optimize the configuration of DRX parameters.

\subsection{DRX Model}

\begin{figure*}
\centering
\includegraphics[scale=0.9]{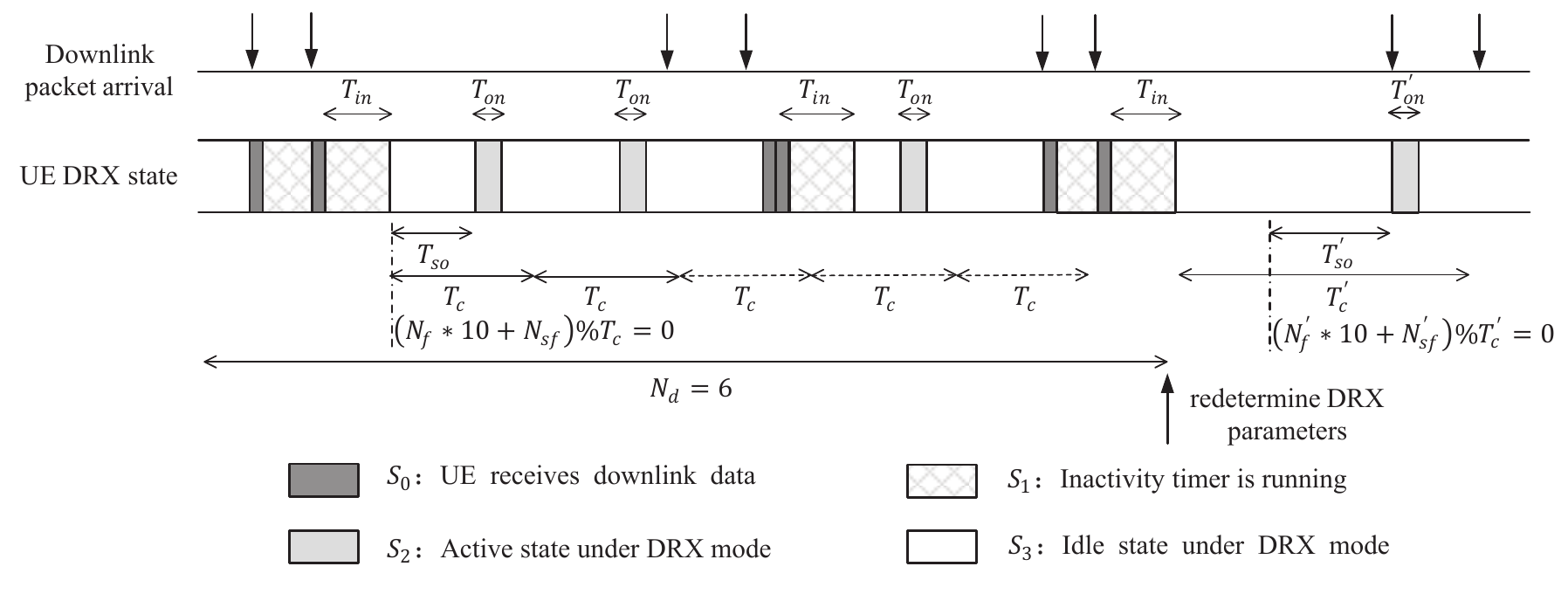}
\caption{UE state transition under DRX mechanism }
\label{DRX model}
\end{figure*}

The DRX mode is an effective way to decrease the power comsumption of the UEs in CIoT. This mode reduces the energy consumed by enforcing the UE go to sleep periodically. 
As mentioned in Section III, the \textit{RadioResourceConfigDedicated} IE is used to realize the configuration of parameters related to the connection, bearer and lower layer functions no matter in the process of establishing, reconfiguring, or re-establishing RRC connections. DRX parameters are included in this IE and transfered to UEs through RRC messages. In other words, the RRC layer configures these DRX parameters and controls the DRX mode of the UEs.

When the BS manages the terminal under CIoT UP scheme, it needs to establish an RRC connection between UE and itself first before it can perform subsequent signaling processes and user data transmission. For UE accessing the network for the first time, UE-specific RRC parameters need to be configured through the RRC connection reconfiguration procedure. Subsequently, UE periodically enters the idle state based on the DRX parameters from the configuration message. If the traffic mode of the UE changes, the RRC layer on the BS side should intelligently inform the UE with the dynamically adjusted DRX parameters through the RRC connection reconfiguration signaling. The DRX parameter configuration algorithm should be based on statistical data of user traffic recorded by the storage module in the RRC layer. The intelligent RRC slicing controller, as the decision-making entity, can take advantage of its historical traffic characteristics to perform service analysis and reconfigure parameters accordingly. The configuration is transferred to the RRC layer through the southbound interface. During this process, the BS continuously monitors and updates the downlink service transmission status of the UE as input information for the algorithm. Then, the slicing control makes decision in time with the execution of the service and enforces RRC layer to adjust the DRX parameters of the UE to fit in the traffic mode.

In Fig. 6, the state transition of the UE under DRX mode is depicted. Under the DRX mechanism of CIoT, UE periodically enters into the idle state, i.e., \textit{RRC\underline{\space}IDLE} state. When the UE is in the connected state, i.e., \textit{RRC\underline{\space}CONNECTED}, downlink data packets that arrive at the BS can be delivered to UE directly, but in the idle state, the data must wait for the end of DRX sleep time and re-enter the active state to monitor the PDCCH channel for downlink control information. Following parameters are mentioned in the above procedures.
\begin{itemize}
\item \textit{Inactivity timer} $T_{in}$: It defines the time that the UE needs to wait after last transmission until entering the DRX mode. Each time the UE successfully receives the downlink data indication from the PDCCH channel, the timer restarts immediately.
\item \textit{DRX cycle} $T_{c}$: DRX cycle defines the period of UE waking up.
\item \textit{On duration timer} $T_{on}$: It is the time during which the UE is in an active state of a DRX period and the UE can monitor the PDCCH to check whether a new packet arrives ($T_{on} \ll T_{c}$). During this time, if the UE can successfully receive the paging message from PDCCH, DRX mode ends at once, and $T_{in}$ starts again.
\item \textit{DRX start offset} $T_{so}$: It is used to define the start time of the on duration time in DRX cycle. When $(10N_{f}+N_{sf}) \bmod T_{c}=T_{so}$ is satisfied, the on duration timer starts running. $N_{f}$ and $N_{sf}$ indicate the current frame number and subframe number respectively.
\end{itemize}

The above parameters are selected by the RRC slicing controller and configured by the RRC layer at the BS side. Usually in networks for human like LTE, the eNodeB configures all UEs with same parameters according to the configuration of the network manager, regardless of UE's traffic pattern. This is alright because humans have similar traffic patterns and energy efficiency is not important compared with the timeliness of messages. While for IoT devices, it is a different situation because devices within one physical network may have various traffic paterns, thus DRX parameters need to be specifically tailored. 

\subsection{Dynamic Control Mechanism}
In order to smartly tune DRX parameters, we resort to the intelligent controller. The decision window length is defined as $N_{d}$ in the unit of downlink arriving data packets during the window to ensure the consistency in size of service features used for analysis in each decision period. At the end of each decision window, the intelligent RRC slicing controller re-determines the DRX parameters of the UE based on the DRX optimization algorithm. If the decision result is different from the current DRX parameter configuration, the slicing controller will inform the RRC layer and the latter will perform RRC reconfiguration process during the first DRX activation period after receiving the command to update the DRX parameters. Otherwise, the current configuration remains unchanged and no action will be taken.

Generally speaking, the UE adopting the DRX mechanism has a total of four different states $S_i, i \in {0,\cdots,3}$, and the corresponding UE powers are $P_i$. In $S_{0}$ state, the UE is receiving the downlink data, and the power consumption is high because transceiver is working. In $S_{1}$ state, the UE waits for data transmission, but is still connected with the BS, and the power consumption is lower than in $S_{0}$ state, but higher than in $S_{2}$ and $S_{3}$ states. The $S_{2}$ and $S_{3}$ states indicate that the UE has entered into DRX mode, and switches between the active and idle state. In the active state, the UE needs to open the receiver to monitor PDCCH channel. While in the idle state, the UE enters deep sleep and only a few modules such as the clock are still working, so the power consumption is extremely low.

Based on the above description, the energy consumption of the UE equals to the sum of energy consumption in each state as shown below.
\begin{equation}
    E_{total}=\sum\nolimits_{i=1}^{4}P_{i} \cdot t_{i}.
\end{equation}
where $t_{i}$ is the time spent on $S_{i}$ state.

The power consumption of the communication module in different states varies depending on the access type and device. Generally, for a device, the longer the total time the UE is in the $S_{3}$ state(i.e., idle deep sleep state), the lower its energy consumption is. Considering the generality of the algorithm applied to different types of CIoT devices, and the lack of reliable measured power data of UE devices, the UE's energy efficiency measurement indicator is simplified as the ratio of time taken by the UE in the DRX sleep state to the total time of the decision interval:
\begin{equation}
    \alpha=\frac{T_{\text{offset}}+N_{c}(T_{c}-T_{on})}{T_{d}},
\end{equation}
where $T_{d}$ is the time interval for making DRX decisions, which is related to the arrival of the downlink data; $T_{\text{offset}}$ indicates the length of the first DRX sleep phase in the decision interval; $N_{c}$ is the number of DRX cycles in the interval.

The average transmission delay of UE data packets is given as:
\begin{equation}
    \beta=\frac{\sum\nolimits_{j=1}^{N_{d}}\Delta t^{j}}{N_{d}},
\end{equation}
where the transmission delay of the $j$-th downlink data packet $\Delta t^{j}$ is the time difference between the arrival and actual transmission of the data packet, i.e. $\Delta t^{j}=t_{tx}^{j}-t_{arrive}^{j}$. Compared to the DRX cycle, the transmission processing time is very short and can be ignored, so we assume that when the UE is in the $S_{0}$, $S_{1}$, $S_{2}$ state, the downlink transmission delay equels zero.

According to the specification, the DRX parameters are all discrete values. There is a trade-off relationship between energy efficiency and downlink transmission delay. If the DRX cycle is small and the UE frequently listens to the channel, the downlink data transmission delay is small but the average power of the UE is high. In contrast, if the DRX cycle is long, the UE is dormant for a long time and the average power is low, but the downlink data transmission delay will increase.
Meanwhile, in the actual IoT application scenarios, different types of services have different requirements for latency and power consumption. For example, low-frequency data collection services such as water metering have low latency requirements but strict power consumption requirements, and instant services such as traffic guidance signals are time-sensitive due to the timeliness of information.
Therefore, we defines an energy efficiency and delay (ED) indicator for evaluating the energy efficiency and delay of different CIoT UEs. This indicator is defined as:
\begin{equation}
    f_{ED}(\lambda,T_{max})=\lambda\alpha+(1-\lambda)(1-\frac{\beta}{T_{max}}),
\end{equation}
where $\lambda \in [0,1]$ is a parameter used to define the relative importance between power consumption and delay in ED index. $T_{max}$ represents the maximum average delay that the UE can tolerate. When the average delay of the UE exceeds $T_{max}$, the delay part of the ED index has a negative impact on the overall value.
The objective is to find a suitable DRX parameter configuration $DRX_{p}$ for the UE to maximize the ED index while meeting the transmission delay requirements at the same time. $DRX_{p}$ is the configuration set of $(T_{c},T_{on},T_{in},T_{so})$. Thus, it can be summarized as:
\begin{equation}
    DRX_{p}=\mathop{\arg\max}_{DRX_{p}}f_{ED}(\lambda,T_{max}).
\end{equation}

\subsection{DRX Optimization Algorithm based on Q-learning}
The arrival times of all downlink data packets are observable by the BS, and the state of the UE upon arrival is indeed affected by DRX parameters. In other words, different DRX parameters $(T_{c},T_{on},T_{in},T_{so})$ may result in different states of the UE even if downlink services arrive at the same time, and therefore affect the power consumption and delay of the UE. In this way, the selection of certain DRX parameters will generate some rewards by reducing the power consumption, and also affect the state of the UE when the downlink data arrives. The reward and environmental state changes are feedback and used by the intelligent controller to make the decision at the end of the period, which is passed to the collaborative management module for connection reconfiguration when the DRX parameter set changes. This process can be modeled as a Markov decision process (MDP) and can be solved by using algorithms based on reinforcement learning.

As the model involves discrete state and discrete action space, so the MDP can be defined as $M=<\mathcal{S},\mathcal{A},T,R>$, where $\mathcal{S}$, $\mathcal{A}$, $T$ and $R$ are the set of states, actions, transition probabilities, and reward respectively. To be specific, the RRC layer executes the action of RRC reconfiguration $a \in \mathcal{A}$ after the intelligent controller makes the decision and passes DRX parameters through the interface. Then the state will change from $s \in \mathcal{S}$ during the last window to $s^{\prime} \in \mathcal{S}$ of the next window. The controller will also receive a real-time reward $R(s,a)$ to evaluate the chosen action. When each data packet arrives, the UE must be in one of four states $S_{i},i \in \{0,1,2,3\}$, which is defined as $s_{i}^{n}$, $n$ is the  sequence number of the downlink packet in the decision window.
So in our solution, the state $s$ is indeed a series of downlink packets' arrival states, so we define the environment state during $k$th decision window as $s^k=\{s_{i}^1,s_{i}^2,...,s_{i}^{N_{d}}\}^k \in \mathcal{S}$, where $i \in \{0,1,2,3\}$. Based on the environmental state, the intelligent controller makes action decisions to reconfigure the parameter set $DRX_{p}$. According to the TS 36.331 document\cite{ts36331}, the action space is like follows.
\begin{equation}
\begin{split}
{T_c} = 256{n_c}{T_{sf}}, \qquad \qquad \quad
\\ {n_c} \in \{ 1,2,4,6,8,12,16,18,24,30,32,36\},
\end{split}
\end{equation}
where $n_{c}$ denotes the number of groups containing $256$ subframes in a period, $T_{sf}$ denotes the time of a subframe, usually 1ms in CIoT.
\begin{equation}
{T_{on}}{\rm{ = }}{n_{on}}{T_{pp}}, \quad {n_{on}} \in \{ 1,2,3,4,8,16,32\},
\end{equation}
where $n_{on}$ is the numbers of PDCCH periods in on duration timer, $T_{pp}$ represents the length of an PDCCH period. In order to simplify the calculation, this paper regards a PDCCH period as 16ms. 
\begin{equation}
{T_{in}} = {n_{in}}{T_{pp}}, \quad {n_{in}} \in \{ 0,1,2,3,4,8,16,32\},
\end{equation}
where $n_{in}$ is the numbers of PDCCH periods of inactivity timer.
\begin{equation}
{T_{so}} = \frac{{{n_{so}}}}{{256}}{T_c}, \quad {n_{so}} \in \{ 0,1,...,255\},
\end{equation}
where $n_{so}$ is the number of subframes by step of $\frac{T_{c}}{256}$.  

The Q-learning algorithm\cite{qlearning} in reinforcement learning can be used to dynamically configure these parameters to meet the UE's requirements on latency and power consumption as shown in Algorithm 1.

\begin{algorithm} 
\caption{DRX parameters configuration based on Q-learning} 
\label{alg1} 
\begin{algorithmic}[1] 
\REQUIRE Initialize the Q value table, discount factor $\gamma$, learning rate $\alpha$ and DRX parameters $DRX_{p}(T_{c},T_{on},T_{in},T_{so})$. 
\ENSURE A series of DRX parameters. 
\REPEAT 
\STATE Calculate the reward $r$ and state $s^{\prime}$ of the last decision cycle $(t_{n-N_{d}} \sim t_{n})$.
	\IF{in exploit mode}
	\STATE Perform feedback value fading detection.
		\IF{$R(s,a)-r<r_{th}$}
		\STATE switch to explore mode
		\ELSE
		\STATE keep the current parameter configuration unchanged.
		\ENDIF
	\ELSIF{in explore mode}
	\STATE update Q-table: $Q(s,a)\leftarrow Q(s,a)+\alpha[r+\gamma max_{a}Q(s^{\prime},a)-Q(s,a)]$;
	\STATE update the current status: $s\leftarrow s^{\prime}$;
	\STATE use $\varepsilon-greedy$ strategy to select the action $a$ to get the corresponding $T_{c},T_{on}$, and $T_{so}^{k + 1} = \frac{1}{{{N_d}}}\sum\limits_{i = (k - 1){N_d} + 1}^{i = k{N_d}} {(({f_i} + s{f_i})\bmod T_c^{k + 1})/T_c^{k + 1}}$, we get the new parameters $DRX_{p}^{\prime}(T_{c},T_{on},T_{in},T_{so})$;
	\IF{$DRX_{p}^{\prime}(T_{c},T_{on},T_{in},T_{so})\neq DRX_{p}(T_{c},T_{on},T_{in},T_{so})$}
	\STATE the BS performs the RRC connection reconfiguration procedure in the next activation state to reconfigures DRX parameters.
	\ENDIF
	\ENDIF
\UNTIL{UE detaches the current cell.} 
\end{algorithmic} 
\end{algorithm}

The algorithm mentioned can be regarded as the dynamic control problem of DRX parameters. As the four parameters $(T_{c},T_{on},T_{in},T_{so})$ work together, and they have an impact on the ED index of the UE. In order to reduce dimensions of action space and accelerate the rapid convergence of the algorithm in practical application scenarios, the algorithm is processed as follows:
\begin{enumerate}[1)]
\item According to the discrete characteristics of DRX parameters, these parameters are encoded and mapped to the corresponding action space except the start offset $T_{so}$. $T_{so}$ is not a typical discrete variable, and it can be changed in real time, so we simply update it based on the new DRX cycle and the data arrival time in last decision period.
\item We define a measurement criterion that can reflect the energy efficiency and downlink transmission delay of the IoT devices. The algorithm learns continuously based on the goal of maximizing this index.
\item In CIoT network, there are often changes in data patterns caused by human or special environmental changes. Therefore, for cases where the reward has dynamic changes, a feedback value fading detection is added to the algorithm to determine the attenuation of real-time feedback under the same state and action. If a certain degree of decline is detected, it indicates that the UE's traffic model has significantly changed. As a result, the predicted reward value needs to be updated, and exploration mode restarts to learn and adapt to the new traffic pattern.
\end{enumerate}

\section{Experiments and Results}
Due to the lack of integrated protocol stacks for our solution, we use \textit{oaisim}\cite{oaisim} to test the procedures and impacts of RRC slicing. \textit{oaisim} is an emulator in OAI and can be used to establish simulated interactions between the UE and the BS while compliant with communication protocols. We verify the content of protocol messages and the correctness of procedures under the OAI simulation environment, and the performance of the implemented RRC slicing is also presented. Moreover, the dynamic algorithm to configure DRX parameters mentioned above is also evaluated.
\subsection{RRC slicing}
In the solution proposed, RRC slicing is implemented in the OAI software, and the RRC layer can be informed by intelligent RRC slicing controller to create, modify and delete slices. For this purpose, the main process of the RRC layer is implemented as a single process with multi-threads. The program parses a configuration file and executes corresponding functions. It supports the initialization, modification and deletion of RRC slices through configuration files. After the main thread for colaborative management is initialized, other threads can be created to perform tasks such as the BS configuration and different RRC slice threads.

\begin{figure*}
\centering
\includegraphics[scale=0.95]{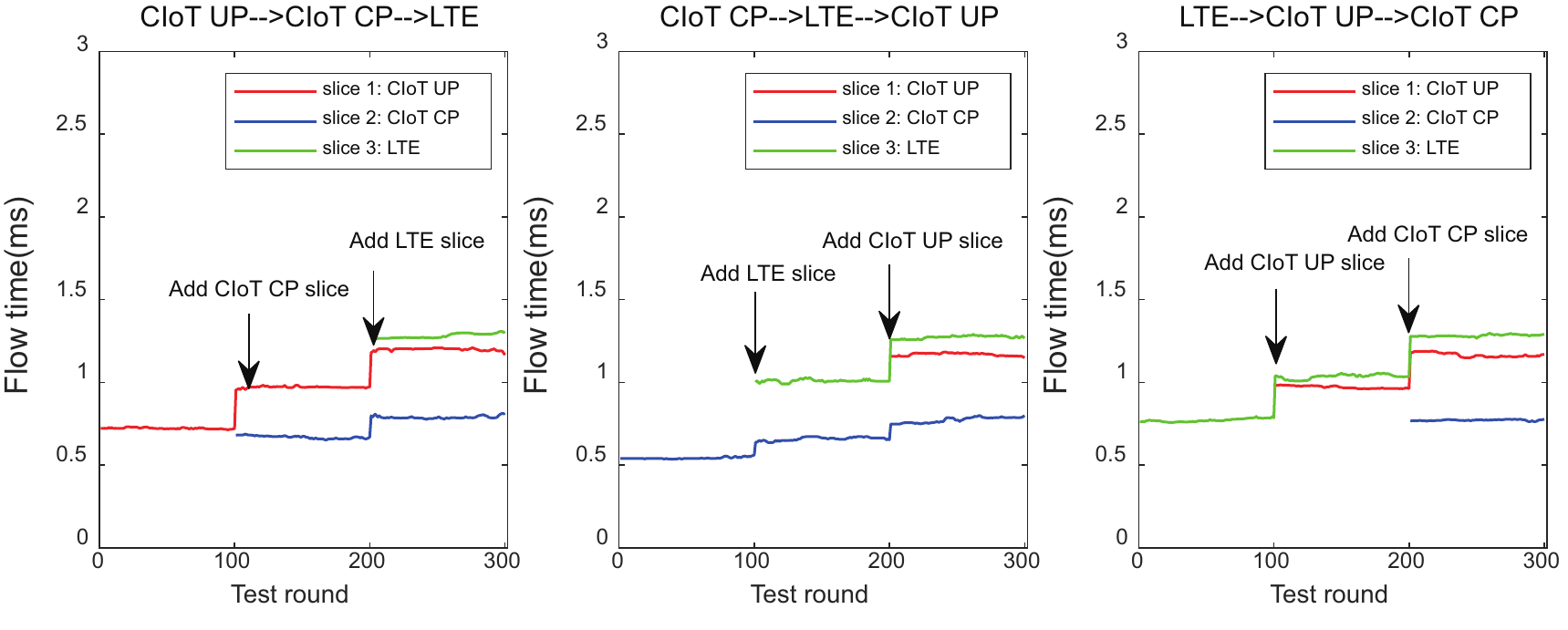}
\caption{Impact of slices on the establishment time of uplink transmission }
\label{6}
\end{figure*}

In our experiment, three IoT scenarios for water metering, bicycle-sharing communication, and video data transmission are chosen to verify the functions of RRC slicing. Three types of RRC slices that meet the requirements of the above scenarios are correspondingly instantiated, namely the NB-IoT with the CIoT CP optimization, and NB-IoT with the CIoT UP optimization and LTE. On one hand, time cost to establish uplink transmission under different number of slices are measured. On the other hand, the impact of adding, modifying or deleting slice on other slices is also validated.
\begin{table*}
	\caption{Impact of slicing operation on the execution time of the RRC procedures (The original number of slices equals 2)}
    \centering
	\renewcommand{\multirowsetup}{\centering}
    \begin{tabular} {|c|c|c|c|c|c|}
\hline
\multirow{2}{3.5cm}{RRC procedures for NB-IoT slice with user plane optimization scheme} & \multicolumn{4}{c|} {\makecell*[c]{average time(ms)}}                        \\ 
\cline{2-5}
                                & \makecell*[c]{regular} & add new slice & modify other slice & delete other slice \\ 
\hline
\makecell*[c]{RRC connection setup}           & 0.563   & 0.616         & 0.623              & 0.584              \\
\hline
\makecell*[c]{AS security estabilishment}      & 0.105   & 0.128         & 0.115              & 0.123              \\
\hline
\makecell*[c]{RRC connection reconfiguration}  & 0.126   & 0.137         & 0.13               & 0.132              \\
\hline
\makecell*[c]{RRC connection suspend}          & 0.041   & 0.063         & 0.047              & 0.059              \\
\hline
\makecell*[c]{RRC connection resume}           & 0.116   & 0.131         & 0.124              & 0.128              \\ 
\hline
\end{tabular}
\end{table*}

As the number of network slices increases, the processing time required to establish an RRC uplink transmission is shown in Fig. 7. Notably, these experiments run on a 4-
core CPU with a main frequency of 3.6GHz. The NB-IoT CP optimization solution can piggyback user data through the NAS message, so the processing time consumed is lower than the other two slices. It can also be seen from the figure that no matter which type of slice, as the number of RRC layer slices increases, the time it takes for a single slice to execute the protocol processing flow will increase slightly. When RRC layer supports 3 network slices, a normal 4-core CPU can guarantee that the processing delay of a one-way flow of each RRC slice does not exceed 200us.

Furthermore, for RRC slices, it is also necessary to meet the needs to minimize the impact on other existing slices when creating, deleting and modifying slice. Table \uppercase\expandafter{\romannumeral2} records the execution time of several RRC procedures of NB-IoT UP slices, which are regular or affected by other slicing operations.
By comparing the execution time of the RRC procedures under regular situation with that in conditions of creating, modifying, or deleting other slices, it can be found that during the processing of the RRC messages, if the BS performs other operations at the same time, it will also slightly affect the current slice in terms of execution time. The cause of this phenomenon is that the execution process is determined by the limited computing resources of the host and the means to allocate and schedule resources between threads of the operating syatem.

It can also be concluded from Table \uppercase\expandafter{\romannumeral2} that modifying and deleting other slices has a smaller impact on the execution flow of the current RRC slice, while creating new slice involves creating a new thread and has a relatively large impact on other slices. As a result, this experiment shows that if the number of CIoT slices maintained by RRC layer is not very large, the design and implementation scheme proposed in this paper can basically guarantee that the operations such as the creation, modification, and deletion of slices will not affect the normal functions of other slices but only slightly affect the process execution time. The impact on execution time can be tolerated in most cases and can be further improved by higher performance processors. However, for service scenarios with ultra low transmission delay requirements, it is necessary to plan and reserve CPU computing resources to meet the requirements of specific delay requirements and satisfy its RRC protocol functions without being affected by other slicing operations.

\subsection{Intelligent controller for DRX}
The previously mentioned DRX dynamic configuration algorithm is implemented in the intelligent RRC slicing controller. In order to verify the actual effect of the algorithm, in addition to the UE and BS simulation interaction functions provided by the OAI platform, the OAI traffic generator (OTG) is also used to produce the downlink traffic for IoT devices. The tool can generate packets with arrival intervals and packet sizes subject to various distributions, e.g. Poisson distribution. In our experiment, the tool can provide simulated data plane transmission for the UE and eNodeB to verify the effect of the algorithm. Under our assumption, the parameters of the ED index are chosen as $\lambda = 0.5$, $T_{max}=0.3$.

In the experiment, we set the patterns of the generated IoT data to change according to certain rules to verify the sensitivity and effectiveness of the algorithm. The inter-data time (IDT) of the CIoT device follows the Poisson distribution and the parameter $\lambda_{\text{IDT}}$ changes dynamically from $\frac{1}{1000}$ to $\frac{1}{6000}$ then to $\frac{1}{200}$. The packet size also follows the Poisson distribution and $\lambda_{\text{Size}}$ changes dynamically from $\frac{1}{600}$ to $\frac{1}{8000}$, then to $\frac{1}{200}$. If we regard the indicator $f_{ED}(0.5,0.3)$ as the optimization target, the result of the DRX cycle output by the dynamic configuration algorithm is shown in Fig. 8, and the results of the DRX on duration timer is shown in Fig. 9.
\begin{figure}
\centering
\includegraphics[scale=0.51]{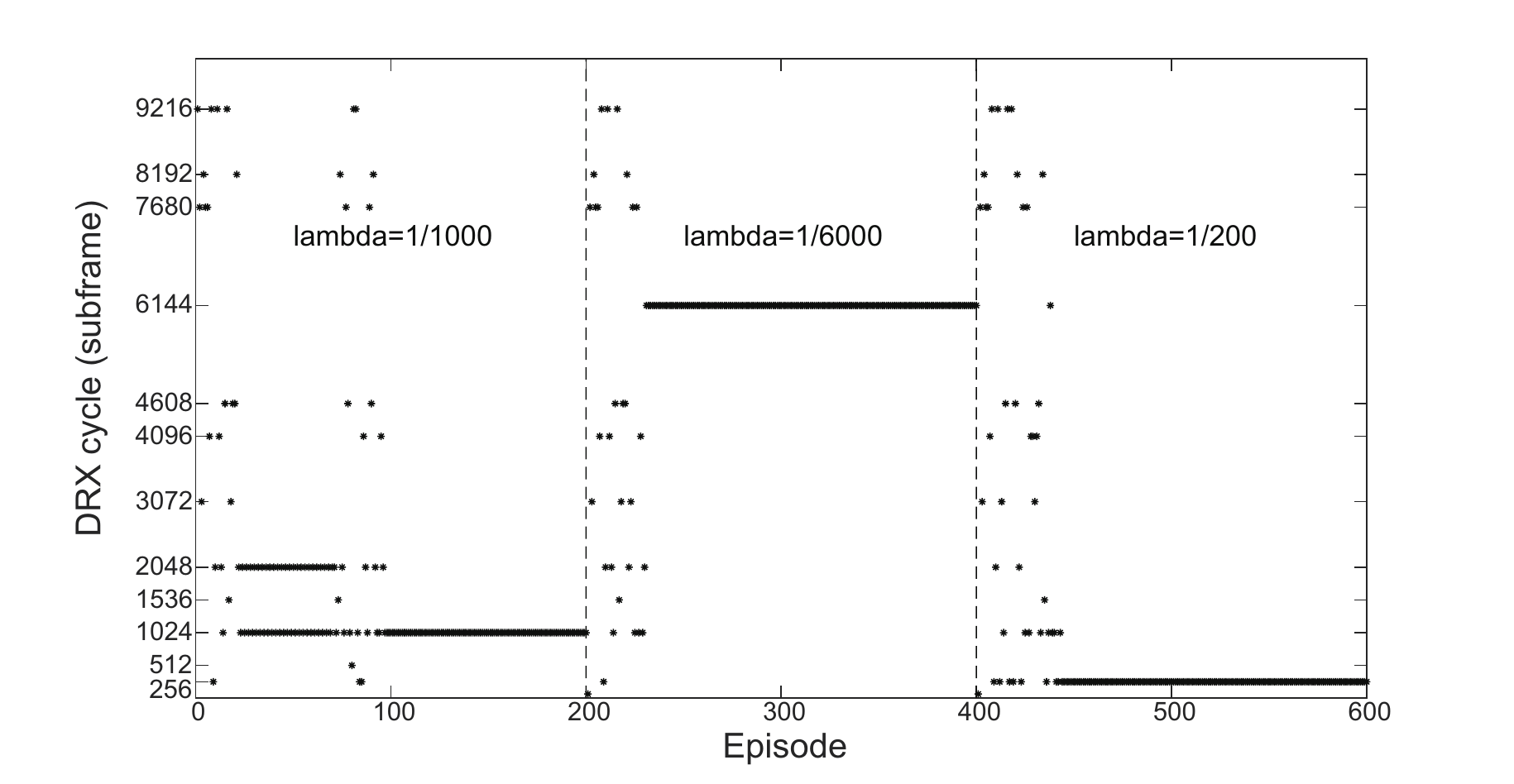}
\caption{Dynamic parameter control for DRX cycle (IDT subjects to Poisson distribution)}
\label{8}
\end{figure}

\begin{figure}
\centering
\includegraphics[scale=0.51]{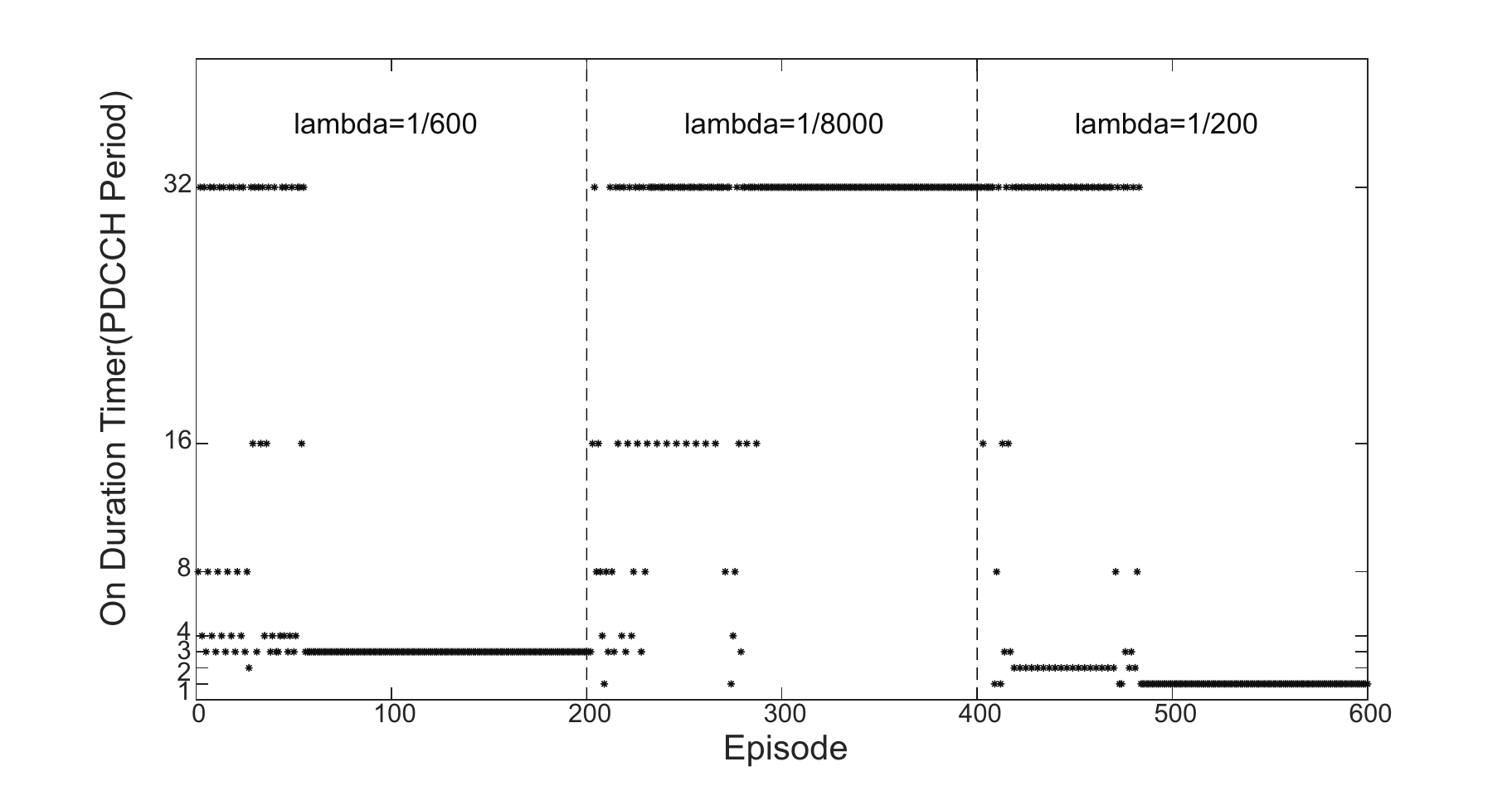}
\caption{Dynamic parameter control for on duration timer (IDT subjects to Poisson distribution)}
\label{9}
\end{figure}

From the results, we can see that the BS is able to adjust the DRX cycle according to the traffic pattern. As the IDT in the UE's traffic model changes, the controller can detect the mismatch of the current DRX cycle from the feedback result, then it re-learns a new DRX cycle. After several rounds of learning (usually no more than 100 iterations) of the decision cycle, the algorithm converges to a new value. Thus the RRC layer of the BS maintains the parameter configuration consistent with the current traffic model. In other words, the optimal configuration is maintained under the stable traffic model. Similarly, the BS can also detect the UE's current DRX on duration timer mismatch in time and re-learn a new value for DRX on duration timer. 

At the same time, the ED index of the UE is shown in Fig. 10. It can be seen that the proposed algorithm based on Q-learning can effectively learn the appropriate DRX parameters to ensure that the ED index value of the UE stay at high level for a long time. Each time the traffic model changes, the process of learning again will make the ED index value fluctuate for a while to explore the suitable parameters.
\begin{figure}
\centering
\includegraphics[scale=0.51]{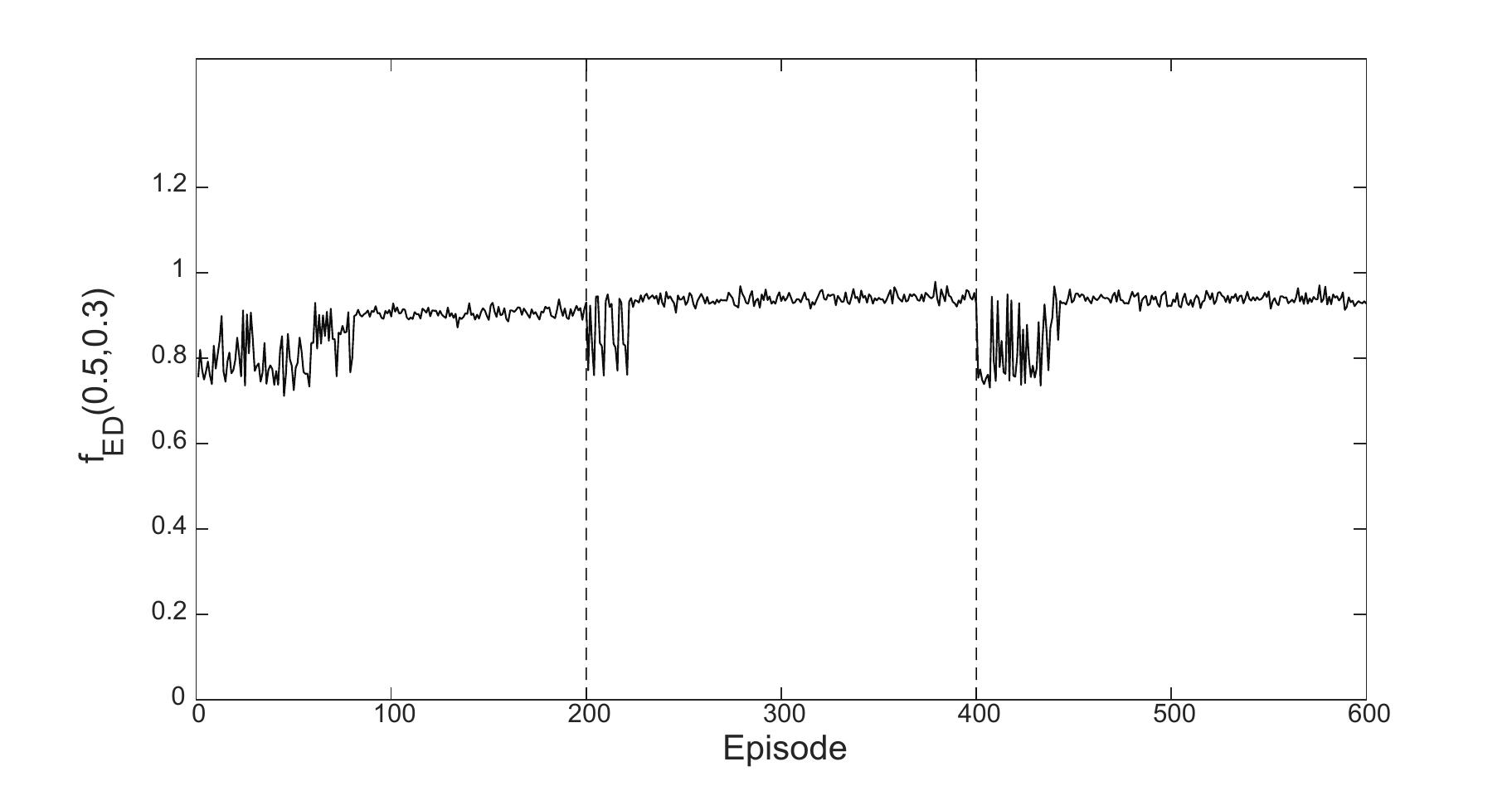}
\caption{Dynamic optimization process for $f_{ED}(0.5,0.3)$ (IDT subjects to Poisson distribution)}
\label{9}
\end{figure}

\section{Conclusions and future challenges}
This paper has presented a CIoT RRC framework to enable the RRC slicing and intelligent control, which is implemented and evaluated in the OAI platform. More specifically, we have presented the implementation methodologies and validated the basic functions of RRC slicing as a proof-of-concept. We have also incorporated an intelligent controller to manage the slices smartly in accordance with the service requirements. The Q-learning based algorithm for DRX parameters configuration is selected as a typical case to manifest the effect of the proposed intelligent RRC slicing architecture. Experiments have shown that the proposed architecture can not only manage differently slices correctly but also customize radio resource control intelligently and dynamically to adapt to the service changes.

We boldly argue that RAN slicing is highly important in the future network and it will change the traditional network architecture that has lasted for decades. Further investigation needs to be conducted in related areas. This calls for not only the evolution in network theory but also efforts from engineering supporters. Moreover, as AI has already been on its way towards maturation, related applications in CIoT systems are an irreversible trend especially in areas of radio resource management and diversified services. In our future work, we will further refine the proposed architecture to better match the practical network configurations and service requirements, and introduce other intelligent algorithms into the system. Similar intelligent algorithms can also be applied in other fields of radio resource control and management. For example, the retransmission times of the physical layer channel in CIoT can be similarly configured based on real-time feedback from the UE, to ensure that UE can adaptively adjust retransmission times according to the channel quality and it's coverage condition. These aspects will be investigated under the framework in the near future.


%

%

%
%

\ifCLASSOPTIONcaptionsoff
  \newpage
\fi




%

\bibliographystyle{IEEEtran}      

\bibliography{IEEEabrv,ref} 
\end{document}